\documentclass[preprint]{sigplanconf}

\usepackage{amsmath}
\usepackage{graphicx}
\usepackage{latexsym}
\usepackage{url}

\usepackage{framed}

\usepackage{color}
\definecolor{darkblue}{rgb}{0.0,0,0.5} 
\definecolor{byzantium}{rgb}{0.44, 0.16, 0.39}
\newcommand{\kw}[1]{{\tt {\color{darkblue} #1}}}

\begin{document}

\setlength{\pdfpageheight}{\paperheight}
\setlength{\pdfpagewidth}{\paperwidth}

\conferenceinfo{CONF 'yy}{Month d--d, 20yy, City, ST, Country} 
\copyrightyear{20yy} 
\copyrightdata{978-1-nnnn-nnnn-n/yy/mm} 
\doi{nnnnnnn.nnnnnnn}



\permissiontopublish             

\preprintfooter{Unpublished, draft paper to be revised. Comments and suggestions are welcome}   

\title{Nez: Practical Open Grammar Language}

\authorinfo{Kimio Kuramitsu}
         {Yokohama National University, JAPAN}
         {kimio@ynu.ac.jp\\http://nez-peg.github.io/}

 
\maketitle

\begin{abstract}

Nez is a PEG(Parsing Expressing Grammar)-based open grammar language that allows us to describe complex syntax constructs without action code. Since open grammars are declarative and free from a host programming language of parsers, software engineering tools and other parser applications can reuse once-defined grammars across programming languages. 

A key challenge to achieve practical open grammars is the expressiveness of syntax constructs and the resulting parser performance, as the traditional action code approach has provided very pragmatic solutions to these two issues. In Nez, we extend the symbol-based state management to recognize context-sensitive language syntax, which often appears in major programming languages. In addition, the Abstract Syntax Tree constructor allows us to make flexible tree structures, including the left-associative pair of trees. Due to these extensions, we have demonstrated that Nez can parse not all but many grammars.

Nez can generate various types of parsers since all Nez operations are independent of a specific parser language. To highlight this feature, we have implemented Nez with dynamic parsing, which allows users to integrate a Nez parser as a parser library that loads a grammar at runtime. To achieve its practical performance, Nez operators are assembled into low-level virtual machine instructions, including automated state modifications when backtracking, transactional controls of AST construction, and efficient memoization in packrat parsing. We demonstrate that Nez dynamic parsers achieve very competitive performance compared to existing efficient parser generators. 

\end{abstract}

\category{D.3.1}{Programming Languages}{Formal Definitions and Theory -- Syntax}
\category{D.3.4}{Programming Languages}{Processors -- Parsing}

\terms
Languages, Algorithms

\keywords
Parsing expression grammars, Context-sensitive grammars, Packrat parsing

\section{Introduction}

A parser generator is a standard approach to reliable parser development in programming languages and many parser applications. Developers use a formal grammar, such as LALR($k$), LL($k$), or PEG, to specify programming language syntax. Based on the grammar specification, a parser generator tool, such as Yacc\cite{Yacc}, ANTLR3/4\cite{PLDI11_Antlr}, or Rats$!$\cite{PLDI06_Rats}, produces an efficient parser code that can be integrated with the host language of the compiler and interpreter. This generative approach, obviously, enables developers to avoid error-prone coding for lexical and syntactic analysis. 

Traditional parser generators, however, are not entirely free from coding. One particular reason is that the formal grammars used today lack sufficient expressiveness for many of the popular programming language syntaxes, such as typedef in C/C++ and other context-sensitive syntaxes\cite{POPL04_PEG,PLDI06_Rats,ONWARD15_Iguana}. In addition, a formal grammar itself is a syntactic specification of the input language, not a schematic specification for tree representations that are needed in semantic analysis. To make up for these limitations, most parser generators take an ad hoc approach called {\em semantic action}, a fragment of code embedded in a formal grammar specification. The embedded code is written in a host language and combined in a way that it is hooked at a parsing context that requires extended recognition or tree constructions. 

The problem with arbitrary action code is that they decrease the reusability of the grammar specification\cite{ICPC08_SemanticActions}. For example, Yacc cannot generate the Java version of a parser simply because Yacc uses C-based action code. There are many Yacc ports to other languages, but they do not reuse the original Yacc grammar without porting of action code. This is an undesirable situation because the well-defined grammar is demanded everywhere in IDEs and other software engineering tools\cite{OOPSLA11_Spoofax}. 

Nez is a grammar language designed for open grammars. In other words, our aim is that once we write a grammar specification, anyone can use the grammar in any programming language. To achieve the openness of the grammar specification, Nez eliminates any action code from the beginning, and provides a declarative but small set of operators that allow the context-sensitive parsing and flexible Abstract Syntax Tree construction that are needed for popular programming language syntaxes. This paper presents the implementation status of Nez with the experience of our grammar development.

The Nez grammar language is based on {\em parsing expression grammars (PEGs)}, a popular syntactic foundation formalized by Ford\cite{POPL04_PEG}. PEGs are simple and portable and have many desirable properties for defining modern programming language syntax, but they still have several limitations in terms of defining popular programming languages. Typical limitations include {\tt typedef}-defined name in C/C++ \cite{POPL04_PEG,PLDI06_Rats}, delimiting identifiers (such as \verb|<<END|) of the Here document in Perl and Ruby, and indentation-based code layout in Python and Haskell \cite{POPL13_Indentation}. 

In the first contribution of Nez, we model a single extended parser state, called a {\em symbol table}, to describe a variety of context-sensitive patterns appearing in programming language constructs. 
The key idea behind the symbol table is simple. We allow any parsed substrings to be handled as {\em symbols}, a specialized word in another context. The symbol table is a runtime, growing, and recursive storage for such symbols. If we handle typedef-defined names as symbols for example, we can realize different parsing behavior through referencing the symbol table. 
Nez provides a small set of symbol operators, including matching, containment testing, and scoping. 
More importantly, since the symbol table is a single, unified data structure for managing various types of symbols, we can easily trace state changes and then automate state management when backtracking. In addition, the symbol table itself is simply a stack-based data structure;  we can translate it into any programming language. 

Another important role of the traditional action code is the construction of ASTs. Basically, PEGs are just a syntactic specification, not a schematic specification for tree structures that represent AST nodes. In particular, it is hard to derive the left-associative pair of trees due to the limited left recursion in PEGs. In Nez, we introduce an operator to specify tree structures, modeled on capturing in perl compatible regular expressions (PCREs), and extend tree manipulations including tagging for typing a tree, labeling for sub-nodes, and left-folding for the left-associative structure. Due to these extensions, a Nez parser produces arbitrary representations of ASTs, which would leads to less impedance mismatching in semantic analysis.
 
To evaluate the expressiveness of Nez, we have performed extensive case studies to specify various popular programming languages, including C, C\#, Java8, JavaScript, Python, Ruby, and Lua. Surprisingly, the introduction of a simple symbol table improves the declarative expressiveness of PEGs in a way that Nez can recognize almost all of our studied languages and then produce a practical representation of ASTs.  
Our case studies are not complete, but they indicate that the Nez approach is promising for a practical grammar specification without any action code. 

At last but not least, parsing performance is an important factor since the acceptance of a parser generator relies definitively on practical performance. In this light, Nez can generate three types of differently implemented parsers, based on the traditional source generation, the grammar translation, and the parsing machine\cite{LPeg}. From the perspective of open grammars, we highlight the latter parsing machine that enables a PCRE-style parser library that loads a grammar at runtime. To achieve practical performance,
Nez operators are assembled into low-level virtual instructions, including automated state modifications when backtracking, transactional controls of AST construction, and efficient memoization in packrat parsing. We will demonstrate that the resulting parsers are portable and achieve very competitive performance compared to other existing standard parsers for Java, JavaScript, and XML. 

The remainder of this paper proceeds as follows. 
Section \ref{sec:nez} is an informal introduction to the Nez grammar specification language. 
Section \ref{sec:design} is a formal definition of Nez. 
Section \ref{sec:condition} describes eliminating parsing conditions from Nez.
Section \ref{sec:impl} describes parser runtime and implementation.
Section \ref{sec:casestudies} reports a summary of our case studies.
Section \ref{sec:perf} studies the parser performance.
Section \ref{sec:relatedwork} briefly reviews related work.
Section \ref{sec:conclusion} concludes the paper.
The tools and grammars presented in this paper 
are available online at \url{http://nez-peg.github.io/}

\section{A Taste of Nez} \label{sec:nez}

This section is an informal introduction to the Nez grammar specification language.

\begin{table}[bt]
\begin{center}
\begin{tabular}{llll} \hline
PEG  & Type & Proc. & Description\\ \hline
\verb|' '| & Primary & 5 & Matches text\\
$[ ]$ & Primary & 5 & Matches character class \\
$.$ & Primary & 5 & Any character\\
$A$ & Primary & 5 & Non-terminal application\\
$( e )$ & Primary & 5 & Grouping\\
$e?$ & Unary suffix & 4 & Option\\
$e*$ & Unary suffix & 4 & Zero-or-more repetitions\\
$e+$ & Unary suffix & 4 & One-or-more repetitions\\
$\&e$ & Unary prefix & 3 & And-predicate\\
$!e$ & Unary prefix & 3 & Negation\\
$e_1 e_2$ & Binary & 2 & Sequencing\\
$e_1 / e_2$ & Binary & 1 & Prioritized Choice\\ \hline

AST  &  &  & \\ \hline

$\{~e~\}$ & Construction & 5 & Constructor\\
$\$(e)$ & Construction & 5 & Connector \\
$\{\$~e~\}$ & Construction & 5 & Left-folding\\ 
$\#t$ & Construction & 5 & Tagging \\
\verb|` `| & Construction & 5 & Replaces a string \\ \hline

Symbols  &  &  & \\ \hline

$\verb|<symbol| ~A\verb|>|$ & Action & 5 & Symbolize Nonterminal $A$ \\ 
$\verb|<exists| ~A \verb|>|$ & Predicate & 5 & Exists symbols \\ 
$\verb|<exists| ~A~x\verb|>|$ & Predicate & 5 & Exists $x$ symbol \\ 
$\verb|<match| ~A\verb|>|$ & Predicate & 5 & Matches symbol \\ 
$\verb|<is| ~A\verb|>|$ & Predicate & 5 & Equals symbol \\ 
$\verb|<isa| ~A\verb|>|$ & Predicate & 5 & Contains symbol \\ 
$\verb|<block| ~e\verb|>|$ & Action & 5 & Nested scope for $e$ \\ 
$\verb|<local| ~A~e\verb|>|$ & Action & 5 & Isolated local scope for $e$ \\ \hline 

Conditional  &  &  & \\ \hline

$\verb|<on| ~c~e\verb|>|$ & Action & 5 & $e$ on c is defined \\ 
$\verb|<on| ~!c~e\verb|>|$ & Action & 5 & $e$ on c is undefined \\ 
$\verb|<if| ~c\verb|>|$ & Predicate & 5 & If c is defined  \\ 
$\verb|<if| ~!c\verb|>|$ & Predicate & 5 & If c is undefined  \\ \hline
\end{tabular}

\caption{Nez operators: "Action" stands for symbol mutators and "predicate" stands for symbol predicates. } 
\label{table:nez}

\end{center}
\end{table}

\subsection{Nez and Parsing Expression}

Nez is a PEG-based grammar specification language. The basic constructs come from those of PEGs, such as production rules and parsing expressions. A Nez grammar is a set of production rules, each of which is defined by a mapping from a nonterminal $A$ to a parsing expression $e$:

\[
A = e
\]

Table \ref{table:nez} shows a list of Nez operators that constitute the parsing expressions. All PEG operators inherit the formal interpretation of PEGs\cite{POPL04_PEG}. That is, the string \verb|'abc'| exactly matches the same input, while \verb|[abc]| matches one of these characters. The . operator matches any single character. The $e?$, $e*$, and $e+$ expressions behave as in common regular expressions, except that they are greedy and match until the longest position. The $e_1\;e_2$ attempts two expressions $e_1$ and $e_2$ sequentially, backtracking to the starting position if either expression fails.  The choice $e_1\;/\; e_2$ first attempt $e_1$ and then attempt $e_2$ if $e_1$ fails. The expression $\&e$ attempts $e$ without any character consuming. The expression $!e$ fails if $e$ succeeds, but fails if $e$ succeeds. Furthermore information on PEG operators is detailed in \cite{POPL04_PEG}.

The expressiveness of PEGs is almost similar to that of {\em deterministic} context-free grammars (such as LALR and LL grammars). In general, PEGs are said to express all LALR grammar languages, which are widely used in a standard parser generator such as Lex/Yacc\cite{Yacc}. In addition, PEGs have many desirable properties that are characterized by:

\begin{itemize}
\item {\em deterministic behavior}, avoiding the dangling if-else problem
\item {\em left recursion-free}, preserving operator precedence
\item {\em scanner-less parsing}, avoiding the extensive use of lexer hacks such as in C++ grammars, and 
\item {\em unlimited lookahead}, recognizing highly nested structures such as  $\{a_n$ $b_n$ $c_n$ $|$ $n$ $>$ $0\}$. 
\end{itemize}

Nez inherits all these properties from PEGs, and has extended features based on AST operators, symbol operators and conditional parsing, as listed in Table \ref{table:nez}. 

\subsection{AST Operators}

The first extension of parsing expressions in Nez is a flexible construction of AST representations. Each node of an AST contains a substring extracted from the input characters. Nez has adopted an PCRE-like capturing operator, denoted by $\{ \}$, to capture the substring. The productions {\tt Int} and {\tt Long} below are examples of capturing a sequence of digits (as defined in {\tt NUM}.)

\begin{verbatim}
  NUM = [0-9]+
  Int = { NUM }
  Long = { NUM } [Ll]
\end{verbatim}

A string to be captured is the exact one matched with {\tt NUM}. That is, {\tt Long} accepts the input \verb|0L| but captures the only substring \verb|0|. An empty string may be captured by an empty expression. 

Tagging is introduced to distinguish the type of nodes. We can just add a \verb|#|-prefixed tag to each of the nodes. 

\begin{verbatim}
  Int = { NUM #Int }
  Long = { NUM #Long } [Ll]
\end{verbatim}

A backquoat operator \verb|` `| is a string operator that replaces the captured string with the specified string. Using the empty capture, we are allowed to create an arbitrary node at any point of a parsing expression.  

\begin{verbatim}
  DefaultValue = { `0` #Int }
\end{verbatim}

The connector \verb|$(e)| is used to make a tree structure by connecting two AST nodes in a parent-child relationship. The prefix \verb|$| is used to specify a child node and append it to a node that is constructed on  the left-hand side of \verb|$(e)|.  As a result, the tree is constructed as the natural order of left-to-right and top-down parsing. 

Basically, the AST operators can transform a sequence of nodes into a tree-formed structure. We can specify that a subtree is nested, flattened, or ignored (by dropping the connector). In the following, we make the construction of a flattened list \verb|(#Add 1 2 3)| and a nested right-associative pair \verb|(#Add 1 (#Add 2 3))| for the same input \verb|1+2+3|.

\begin{verbatim}
  List = { $(Int) ('+' $(int))* #Add }
  Binary = { $(Int) ('+' $(Binary)) #Add }
\end{verbatim}

The construction of the left-associative structure, however, suffers from the forbidden left-recursion (e.g, {\tt Binary = Binary \verb|'+'| Int}). Nez specifically  provides a left-folding constructor \verb|{$ ...}| that allows a left-hand-constructed node to be contained in a new right-hand node. Accordingly, we can make a precedence-preserving form of ASTs by folding from the repetition. 

\begin{verbatim}
  Add = Int {$ '+' $(int) #Add}* 
\end{verbatim}

The connector and the left-folding operator can associate an optional label for the connected child node. The associated label follows the \verb|$| notion:

\begin{verbatim}
  Add = Int {$left '+' $right(int) #Add}* 
\end{verbatim}

Note that the AST representation in Nez is a so-called common tree, a sort of generalized data structure. However, we can map a tag to a class and then map labels to fields in a mapped class, which leads to automated conversion of concrete tree objects. Although mapping and type checking are an interesting challenge, they are beyond the scope of this paper. 

\begin{figure}[tb]
\begin{small}
\begin{verbatim}
Expr = Prod {$left ('+' #Add / '-' #Sub) $right(Prod)}*
Prod = Val {$left ('*' #Mul / '/' #Div) $right(Val)}*
Val = { [0-9]+ #Int }
\end{verbatim}
\end{small}
\begin{center}
\includegraphics[width=7.0cm]{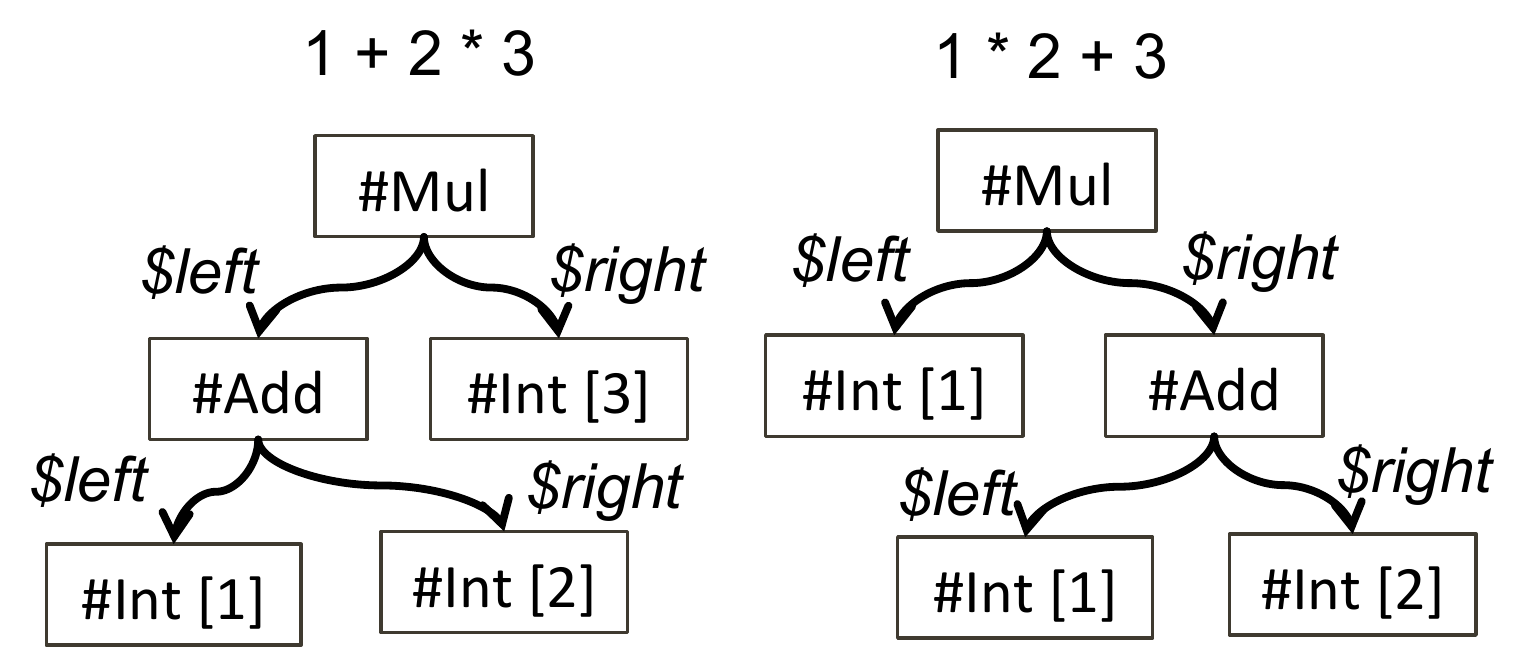}
\end{center}
\caption{Mathematical basic operators and example of ASTs}
\label{fig:math}
\end{figure}

To the end, we show an example of mathematical basic operators written in Nez. Figure \ref{fig:math} shows some of ASTs constructed when evaluate the nonterminal {\tt Expr}.  

\subsection{Symbol Table}

PEGs are very expressive, but they sometimes suffer from insufficient expressiveness for context-sensitive syntax, appearing even in popular programming languages such as: 

\begin{itemize}
\item Typedef-defined name in C/C++ \cite{POPL04_PEG,PLDI06_Rats}
\item Here document in Perl, Ruby, and many other scripting languages
\item Indentation-based code layout in Python and Haskell \cite{POPL13_Indentation}
\item Contextual keywords used in C\# and other evolving languages\cite{OOPSLA06_AspectJ}
\end{itemize}

The problem with context-sensitive syntax is that PEGs are assumed stateless parsing and cannot handle state changes in the parser context. However, most of the state changes above can be modeled by a {\em string specialization};  the meaning of words is changed in a certain context. We call such a specialized string a {\em symbol}. Nez is designed to provide a symbol table and related symbol operators to manage symbols in the parser context.

To illustrate what a symbol is, let us suppose that the production {\tt NAME} is defined:

\begin{verbatim}
 NAME = [A-Za-z] [A-Za-z0-9]*
\end{verbatim}

The symbolization operator takes the form of \verb|<symbol |$A$\verb|>| to declare that a substring matched at the nonterminal $A$ is a symbol. We call such a symbol an $A$-{\em specialized symbol}, which is stored with the association of the nonterminal in the symbol table. 

Here is a symbolization of {\tt NAME}, where  {\tt NAME} first attempts to match and, if matched, {\small \verb|<match NAME>|} adds the matched symbol to the symbol table. 

\begin{verbatim}
 <symbol NAME>
\end{verbatim}

In Nez, a {\em symbol predicate} is defined to refer to the symbol by the nonterminal name. The following {\small \verb|<match NAME>|} is one of the symbol predicates that exactly match the \verb|NAME|-specialized symbol for the input characters. 

\begin{verbatim}
 <match NAME>
\end{verbatim}

Compared to the nonterminal {\tt NAME}, the result of \verb|<match NAME>| varies, depending on the past result at \verb|<symbol NAME>|. That is, if the \verb|<symbol NAME>| accepts {\tt Apple} previously, \verb|<match NAME>| only accepts \verb|'Apple'|. In this way, the symbol operators handle context-depended parsing with the state changes in the symbol table.

The uniqueness of Nez is that the state changes are limited to a single symbol table. That is, we are allowed to add multiple symbols with \verb|<symbol NAME>| or different kinds of symbols with another nonterminal in the same table. Nez offers various types of symbol predicates, which can apply different patterns of the symbol references.

\begin{itemize}
\item {\small $\verb|<exists|~A \verb|>|$} -- checks if the symbol table has any $A$-specialized symbols
\item {\small $\verb|<exists|~A~s \verb|>|$} -- checks if the symbol table has an $A$-specialized symbol that is equal to the given string $s$ 
\item {\small $\verb|<match|~A \verb|>|$} -- matches the last $A$-specialized symbol over the input characters
\item {\small $\verb|<is|~A \verb|>|$} -- compares the last $A$-specialized symbol with an $A$-matched substring.
\item {\small $\verb|<isa|~A \verb|>|$} -- checks the containment of an $A$-matched substring in a set of $A$-specialized symbols stored in the table. 
\end{itemize}

Note that the two symbol predicates {\small $\verb|<match|~A \verb|>|$} and {\small $\verb|<is|~A \verb|>|$} are quite similar, but they significantly differ in terms of the input acceptance.
To illustrate the difference, suppose that the previous {\small \verb|<match NAME>|} accepts and stores the symbol \verb|'in'|. {\small \verb|<match NAME>|} accepts the input \verb|'include'| (i.e., the input \verb|'clude'| is unconsumed), while {\small \verb|<is NAME>|} does not because it compares the stored symbol \verb|'in'| with the {\tt NAME}-matched string \verb|'include'|. 

Figure \ref{fig:typedef} shows a simplified example of the typedef-name syntax, excerpted from the C grammar. 
The production {\tt TypeDef} describes the syntax of the {\tt typedef} statement, which includes the symbolization of {\tt USERTYPE}. In {\tt TypeName}, we first match built-in type names and then match one of the {\tt USERTYPE}-specialized symbols. As a result, we can express context-sensitive patterns in {\tt TypeName}.

\begin{figure}[tb]

{\small 
\begin{framed}
\begin{verbatim}
USERTYPE  = [A-Za-z_] !W*
W = [A-ZA-z_0-9]
S  = [ \t\r\n]
  
TypeDef  
  = 'typedef' S* TypeName S* <symbol USERNAME> S* ';'

TypeName 
  = BuildInTypeName / <isa USERNAME>

BuiltInType
  = 'int' !W / 'long' !W / 'float' !W ...

\end{verbatim} 
\end{framed}
}
\caption{Definition of typedef-name syntax with symbol operators}
\label{fig:typedef}

\end{figure}

\subsubsection{Scoping}

In principle, symbols in the symbol table are globally referable from any production. However, many programming languages have their own scope rule for identifiers, which may require us to restrict the scope of symbols in parallel with a scope construct of the language.  Nez provides the explicit scope declaration for that purpose. 

Here, we consider a simple case where XML tags need to match closed tags with the same open tags.

{\small \begin{verbatim}
  <A><B> ... </B> </A>
\end{verbatim}}

Briefly, an XML element can be specified as follows: 

{\small \begin{verbatim}
  ELEMENT = 
           <symbol TAG> ELEMENT <is TAG>
\end{verbatim}}

As described above, the symbol {\tt TAG} enables us to ensure the same name in both open and closed tags. However, the {\tt ELEMENT} involves nested symbolization inside, resulting in repeated symbolization. On the other hand, {\small \verb|<is TAG>|} refers to the latest {\tt TAG}-specialized symbol. As a result, the {\tt ELEMENT} above can only accept the last tag, such as {\small \verb|<A><B> ... </B> </B>|}, which is a not desirable result. 

The notation {\small $\verb|<block|~e\verb|>|$} is used to declare a nested scope of symbols. That is, any symbols defined inside {\small $\verb|<block|~e\verb|>|$} are {\em not} referable outside the block. Here is a scoped version of the ELEMENT, which works as expected

{\small \begin{verbatim}
  ELEMENT = <block 
           <symbol TAG> ELEMENT <is TAG> >
\end{verbatim}}


In Nez, we have adopted a single nested scope for multiple nonterminal symbols. In other words, different kinds of symbols are equally controlled with a single scope. One could consider that there is a language that has a more complex scoping rule, while our focus is only on names that directly influence the syntactic analysis. One important exception is an explicit isolation of the specific symbols by {\small $\verb|<local|~A~e~\verb|>|$}. In this scope, all $A$-specialized symbols before the local scope are isolated and not referable by the subexpression $e$.

\subsection{Conditional Parsing}

The idea of conditional parsing is inspired by the {\em conditional compilation}, where the {\tt \#ifdef ... \#endif} directives switch the compilation behavior. Likewise, Nez uses \verb|<if ...>| to switch the paring behavior, but it differs in that the condition, as well as the semantic predicate, can be controlled by \verb|<on ...>| in the parser context. 

Nez supports multiple parsing conditions, identified by the user-specified condition name. Let $c$ be a condition name. A parsing expression $\verb|<if| ~c\verb|>| ~ e$ means that $e$ is evaluated only if $c$ is {\em true}. We allow the $!$ predicate for the negation of the condition $c$. That is, $\verb|<if| ~!c\verb|>|$ is a syntax sugar of $!\verb|<if| ~c\verb|>|$. Two expressions $e_1$ and $e_2$ are distinctly switched by $\verb|<if|~c\verb|>| ~ e_1 ~/~ \verb|<if|~!c\verb|>| ~ e2$. 
 
In addition, Nez provides the scoped-condition controller for arbitrary parsing expressions. $\verb|<on| ~c~e\verb|>|$ means that the subexpression $e$ is evaluated under the condition that $c$ is true, while $\verb|<on| ~!c~e\verb|>|$ means that the subexpression $e$ is evaluated under the condition that $c$ is false. 
 
Here is an example of conditional parsing; the acceptance of  {\tt Spacing} below depends on the condition {\tt IgnoreNewLine}:

{\small \begin{verbatim}
Spacing = <if !IgnoreNewLine> [\n\r] / [ \t]
\end{verbatim}}

The conditions, as well as the symbols, are a global state across productions. That is, the condition {\tt IgnoreNewLine} affects all nonterminals and parsing expressions that involve the {\tt Spacing} nonterminal. The expression $\verb|<on| ~{\tt IgnoreNewLine} ~e\verb|>|$ is used to declare the static condition of the inner subexpression. The following are the {\tt IgnoreNewLine} version and non-{\tt IgnoreNewLine} version of the {\tt Expr} nonterminal. 

{\small \begin{verbatim}
<on IgnoreNewLine Expr>
<on !IgnoreNewLine Expr>
\end{verbatim}}

Note that conditional parsing is a Boolean version of the symbol table. Let $C$ be an empty expression (i.e., $C = \verb|''|$).  {\small $\verb|<on|~c~e~\verb|>|$} is a syntax sugar of {\small $\verb|<block <symbol|~C~ \verb|>|~e~ \verb|>|$} and {\small $\verb|<if|~c~\verb|>|$} is a syntax sugar of  {\small $\verb|<exists|~C~\verb|''>|$}. However, the conditional parsing can be eliminated from arbitrary Nez grammars, since the possible states are at most 2. This is why Nez provides specialized operators for conditional parsing. The elimination of parsing conditions is described in Section \ref{sec:condition}.

\section{Grammar and Language Design} \label{sec:design}

This section describes the formal definition of Nez. Figure \ref{fig:notation} is a list of notations used throughout this section.

\begin{figure}
\begin{tabular}{ll}
$A,B,C \in N$ & Nonterminals
\\
$a,b \in \Sigma$ & Characters
\\
$x,y,z \in \Sigma^{*}$ & Sequence of characters
\\
$xy$ & Concatenation of x and y
\\
$e$ & Parsing expressions
\\
$A=e$ & Production 
\\
$T$ & Tree node 
\\
$[A, x]$ & $A$-specialized symbol $x$
\\
$S=[A,x][A', y]$ & State, sequence of labeled symbols
\\
$\epsilon$ & empty string or sequence
\\
\end{tabular}
\caption{Notations uses throughout this paper}
\label{fig:notation}
\end{figure}

\subsection{ASTs and Symbol Table}

\begin{figure}[tb]
\begin{center}
\includegraphics[width=7.0cm]{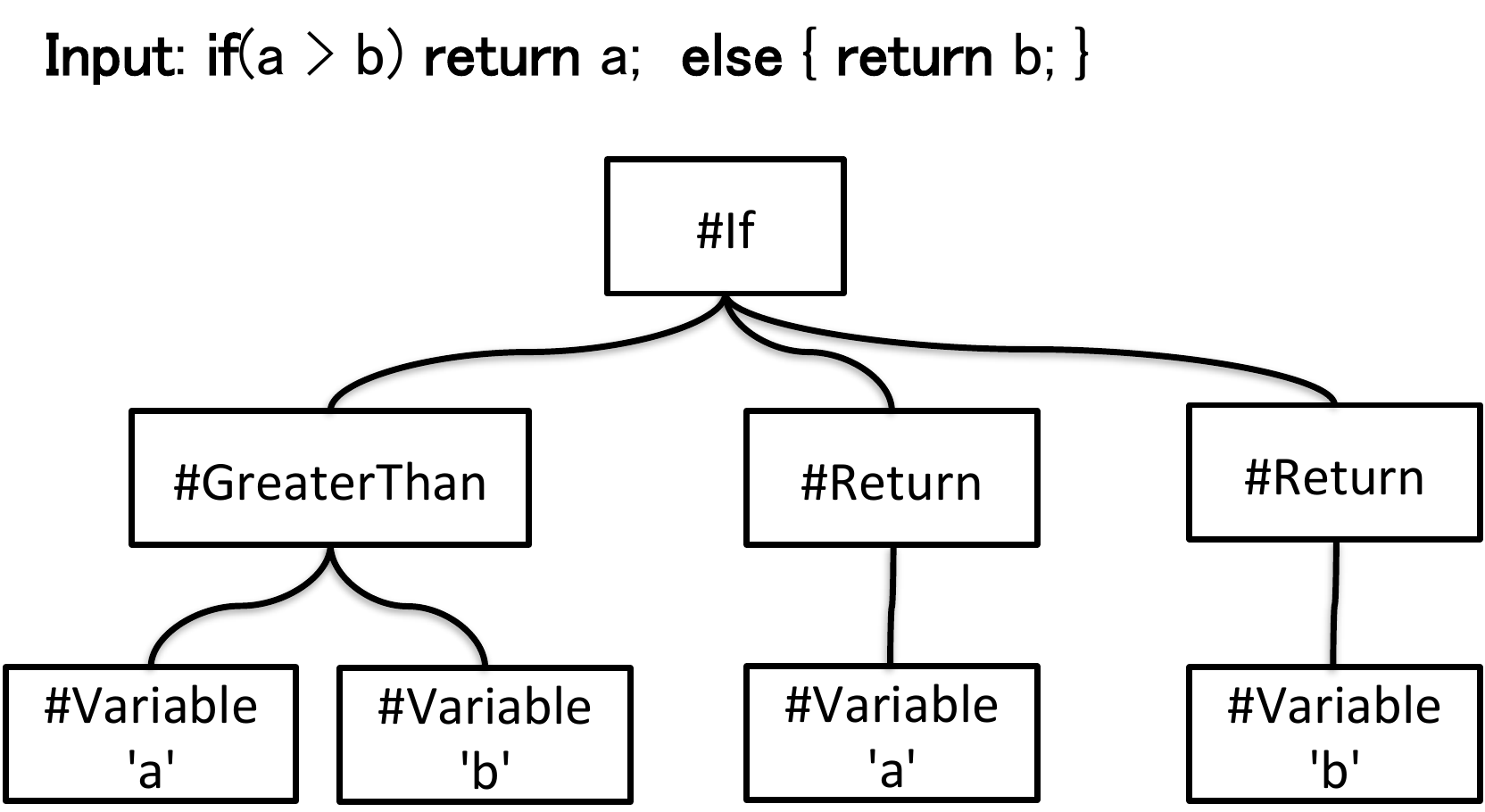}
\end{center}
\caption{Pictorial notation of ASTs}
\label{fig:ast}
\end{figure}

We start by defining the model of the AST representation and the symbol table in Nez.

An AST is a tree representation of the abstract structure of parsed results. The tree is "abstract" in the  sense that it contains no unnecessary information such as white spaces or grouping parentheses. Based on the AST operators in Nez, the syntax of the AST representation, denoted by $v$, is defined inductively:


\[
 v ~ \verb|:==| ~~ \#t[v] ~~~ | ~~~ \#t[\verb|'...'|] ~~~ | ~~~v ~ v
\]

\noindent where $\#t$ is a tag to identify the type of $T$ and a captured string written by \verb|'...'|. A whitespace concatenates two or more nodes as a sequence. 
Note that we ignore the labeling of subnodes for simplicity. 


Here is an example of an AST tree, shown in Figure \ref{fig:ast}.

{\small \begin{verbatim}
  #If[
    #GreaterThan[
      #Variable['a'] 
      #Variable['b']
    ]
    #Return[#Variable['a']]
    #Return[#Variable['b']]
  ] 
\end{verbatim}}

The initial state of the AST node is an empty tree. $\texttt{new}(x)$ is a function that instantiates a new AST node including a given string $x$. Let $v$ be a reference of a node of ASTs. That is, we say $v = v'$ if $v'$ is the mutated $v$ with the following tree manipulation functions: 

\begin{itemize}
\item $\text{tag}(v, \#t)$ -- replaces the tag of $v$ with the specified $\#t$
\item $\texttt{replace}(v, x)$ -- replaces the string of $v$ with the specified string $x$
\item $\text{link}(v, v')$ -- appends a child node $v'$ to the parent node $v$
\end{itemize}

Next, we will turn to the model of the symbol table. Let $[A, x]$ denote a symbol $x$ that is specialized by nonterminal $A$. The symbol table $S$ is represented by a sequence of symbols, such as $[A,x][B,y] ... $. Suppose two following tables:\\

$S_1 = [A,x][B,y]$

$S_2 = [A,x][B,y][A,z]$ \\

The initial state of the symbol table is an empty sequence, denoted by $\epsilon$. We write $[A,x] \in S$ for the containment of $[A,x]$ in $S$. In addition, we write $S [A,z]$ to represent an explicit addition of $[A, z]$ into $S$. ($S S'$ is a concatenation of $S$ and $S'$) That is, we say that $S_1 [A,z] = S_2$ and $S_1 \subset S_2$. Other operations for $S$ are represented by the following two functions:

\begin{itemize}
\item $top(S, A)$ -- a $A$-specialized symbol that is recently added in the table. That is, we say $top(S_1, A) = x$ and $top(S_2, A) = z$
\item $del(S, A)$ -- a new table that removes all $A$-specialized entries from $S$. 
\end{itemize}

\subsection{Grammars}


A Nez $G = (N, \Sigma, P, e_s, \mathcal{T})$ has elements, where $N$ is a set of nonterminals, 
$\Sigma$ is a set of characters, $P$ is a set of productions, $e_s$ is a start expression, and $\mathcal{T}$ is a set of tags.

\begin{figure}[tb]
\begin{center}
\begin{tabular}{lrll} 
$e$ &  \verb|      ::= | & $\epsilon$ & : empty \\ 
& \verb#|  # & $A$ & : nonterminal  \\
& \verb#|  # & a & : character \\
& \verb#|  # & $e ~ e' $ & : sequence \\
& \verb#|  # & $e ~ / ~ e' $ & : prioritized choice \\
& \verb#|  # & $e*$ &  : repetition  \\
& \verb#|  # & \verb|&|$e$ &  : and predicate  \\
& \verb#|  # & \verb|!|$e$ &  : not predicate  \\
& \verb#|  # & $\{~e~\}$ &  : constructor  \\
& \verb#|  # & $\{\$~e~\}$ &  : left-folding  \\
& \verb#|  # & $\$(e)$ &  : connector  \\
& \verb#|  # & $\#t$ &  : tag  \\
& \verb#|  # & $`x`$ &  : string replacement  \\
& \verb#|  # & $\langle symbol~A\rangle$ & : symbol definition \\
& \verb#|  # & $\langle exists~A$ & : symbol existence  \\
& \verb#|  # & $\langle exists~A~x\rangle$ & : symbol existence  \\
& \verb#|  # & $\langle match~A\rangle$ & : symbol match \\
& \verb#|  # & $\langle is~A\rangle$ & : symbol equivalence \\
& \verb#|  # & $\langle isa~A\rangle$ & : symbol containment  \\
& \verb#|  # & $\langle block~e\rangle$ & : nested scope \\
& \verb#|  # & $\langle local~A~e\rangle$ & : isolated local scope  \\
\end{tabular}
\end{center}
\caption{An abstract syntax of Nez expressions}
\label{fig:syntax}
\end{figure}

Each production $p \in P$ is a mapping from $A$ to $e$; we write as $A=e$, where $A \in N$ and $e$ is a parsing expression. Figure \ref{fig:syntax} is an abstract syntax of parsing expressions in Nez. Due to space constraints, we focus on core parsing expressions. Note that  $e?$ is the syntax sugar of $e / \epsilon$, $e*$ the sugar of $A' = e A' / \epsilon$, and $e+$ the sugar of $e e*$ (as defined in \cite{POPL04_PEG}).


The transition form $(S, xy, v) \xrightarrow{e} (S', x, v')$ may be read: In symbol table $S$, the input sequence $xy$ is reduced to characters $y$ while evaluating $e$ (i.e, $e$ consumes $x$). Simultaneously, the semantic value $v$ is evaluated to $v'$.  The symbol table is unchanged if $(S, x, v) \xrightarrow{e} (S S', y, v')$ and $S'$ is empty. Likewise, the AST node is not mutated if  $(S, x, v) \xrightarrow{e} (S', y, v')$ and $v = v'$. 

We write $\bullet$ for a special failure state, suggesting backtracking to the alternative if it exists. Here is an example failure transition. 

\[
     (S, bx, v)  \xrightarrow{a} \bullet    ~~~~~~ (a \ne b)
\]

Figure \ref{fig:sempeg} is the definition of the operational semantics of PEG operators. Due to the space constraint, we omit the failure transition case. Instead, undefined transitions are regarded as the failure transition. The semantics of Nez is conservative and shares with the same interpretation of PEGs. Notably, the and-predicate $\& e$ allows the state transitions of $S \mapsto S S'$ and $v \mapsto v'$, meaning that the lookahead definition of symbols and the duplication of local AST nodes.


\begin{figure}[tb]
\begin{small}
\fbox{$\epsilon$} \vspace{-3mm}
\[
{(S, x, v) \xrightarrow{\epsilon} (S, x, v)}
\]

\fbox{$a$} \vspace{-3mm}
\[
{(S, ax, v) \xrightarrow{a} (S, x, v)}
\]

\fbox{$A = e$} \vspace{-3mm}
\[
\frac{~~ (S, xy, v) \xrightarrow{e} (SS', y, v')}
{(S, xy) \xrightarrow{A} (SS', y, v')}
\]

\fbox{$e~e'$} 
\[
\frac{~~ (S, xyz, v) \xrightarrow{e} (SS', yz, v') ~~~ (SS', yz, v') \xrightarrow{e'} (SS'S'', z, v'')}
{(S, x, v) \xrightarrow{e ~e'} (SS'S'', z, v'') ~~}
\]

\fbox{$e/e'$} \vspace{-3mm}
\[
\frac{~~ (S, xy, v) \xrightarrow{e} (SS', y, v') }{(S, xy, v) \xrightarrow{e/e'} (SS', y, v')}
\]

\[
\frac{~~ (S, xy, v) \xrightarrow{e} \bullet ~~~ (S, xy, v) \xrightarrow{e'} (SS', y, v') }{(S, xy, v) \xrightarrow{e/e'} (SS', y, v')}
\]

\fbox{$\&e$} \vspace{-3mm}
\[
\frac{~~ (S, xy, v) \xrightarrow{e} (SS', y, v')~~ }{(S, xy, v) \xrightarrow{\&e} (SS', xy, v')}
\]

\fbox{$!e$} \vspace{-3mm}
\[
\frac{~~ (S, x, v) \xrightarrow{e} \bullet ~~}{(S, x, v) \xrightarrow{!e} (S, x, v)}
\]
\end{small}
\caption{Semantics of PEG operators}
\label{fig:sempeg}
\end{figure}

\begin{figure}[tbh]
\begin{small}
\fbox{$\{ ~e~ \}$} \vspace{-3mm}
\[
\frac{~~ (S, xy, v) \xrightarrow{e} (SS', y, v')}
{(S, xy, T) \xrightarrow{\{\$~e~\}} (SS', y, \texttt{new}(x))}
\]

\fbox{$\$(e)$} \vspace{-3mm}
\[
\frac{~~ (S, xy, v) \xrightarrow{e} (S', y, v')}
{(S, xy, v) \xrightarrow{\$(e)} (S', y, \texttt{link}(v, v'))}
\]

\fbox{$\{\$ ~e~ \}$} \vspace{-3mm}
\[
\frac{~~ (S, xy, v) \xrightarrow{e} ('S, y, v')}
{(S, xy, v) \xrightarrow{\{\$~e~\}} ('S, y, \texttt{link}(\texttt{new}(x),v)  )}
\]

\fbox{$\#t$} \vspace{-3mm}
\[
      (S, x, v)  \xrightarrow{\#t} (S, x, \texttt{tag}(v, \#t))
\]

\fbox{$`x`$} \vspace{-3mm}
\[
      (S, x, v)  \xrightarrow{`x`} (S, x, \texttt{replace}(v, x))
\]

\end{small}

\caption{Semantics of AST operators}
\label{fig:semast}
\end{figure}

\begin{figure}[bt]
\begin{small}
\fbox{$\langle symbol~A\rangle$} \vspace{-3mm}
\[
\frac{~ (S, xy, T) \xrightarrow{A} (S, y, T')~~}
{(S, xy, T) \xrightarrow{\langle symbol~A\rangle} (S [A,x], y, T')~~}
\]

\fbox{$\langle block~e\rangle$} \vspace{-3mm}
\[
\frac{~~ (S, xy, T) \xrightarrow{e} (S S', y, T')~~}{(S, xy, T) \xrightarrow{\langle block~e\rangle} (S, y, T')}
\]

\fbox{$\langle local~A~e\rangle$}
\[
\frac{~~ S_{\bar{A}} = del(S,A) ~~~ (S_A, xy, T) \xrightarrow{e} (S', y, T') ~~}
{(S, xy, T) \xrightarrow{\langle local~A~e\rangle} (S, y, T)}
\]

\fbox{$\langle exists~A\rangle$} \vspace{-3mm}
\[
\frac{~~ \exists z [A,z] \in S ~~} {(S, x, T) \xrightarrow{\langle exists~A\rangle} (S, x, T)}
\]

\fbox{$\langle exists~A~x\rangle$} \vspace{-3mm}
\[
\frac{~~ \exists [A,z] \in S ~~} 
{(S, x, T) \xrightarrow{\langle exists~A~x\rangle} (S, x,T)}
\]

\fbox{$\langle match~A\rangle$} \vspace{-3mm}
\[
\frac{~~  top(S,A) = z ~~ }{(S, zx, T) \xrightarrow{\langle match~A\rangle} (S, x, T)}
\]

\fbox{$\langle is~A\rangle$} \vspace{-3mm}
\[
\frac{~~ (S, zx, T) \xrightarrow{A} (S, x, T') ~~ top(S,A) = z ~~ }
{(S, zx, T) \xrightarrow{\langle is~A\rangle} (S, x, T')}
\]

\fbox{$\langle isa~A\rangle$} \vspace{-3mm}
\[
\frac{~~ (S, zx, T) \xrightarrow{A} (S, x, T') ~~ \exists [A,z] \in S~~}
{(S, zx, T) \xrightarrow{\langle isa~A \rangle} (S, x, T')}
\]
\end{small}
\caption{Semantics of Nez's symbol operators}
\label{fig:semsym}
\end{figure}

Figure \ref{fig:semast} and Figure \ref{fig:semsym} show the semantics of AST operators and symbol operators. The recognition of Nez is AST-independent, because any values $v$ are not premise for the transitions. Accordingly, an AST-eliminated grammar accepts the same input with the original Nez grammar. 

Formally, the language generated by the expression $e$ is $L(S, e) = \{ x | (S, xy) \xrightarrow{e} (S', y) \}$, and the language of grammar L(G) is $L(G) = \{ x | (\epsilon, xy) \xrightarrow{e_s} (S, y) \}$. Note that the semantic value is unnecessary in the language definition. As suggested in the study of predicated grammars\cite{PLDI11_Antlr}, the language class of $L(G)$ is considered to be {\em context-sensitive} because predicates can check both the left and right context. 

\section{Elimination of Parsing Condition} \label{sec:condition}


In the previous section, we described the formal definition of Nez grammar language without conditional parsing. One reason for this is that conditional parsing can be replaced with symbol operators and an empty symbol.  In addition, we can eliminate conditions from a Nez grammar, and Nez parsers usually run based on the eliminated grammar. This section describes how to eliminate the conditions.

For simplicity, we consider the a single parsing condition, labeled $c$. Let $x$ be a Boolean variable such as $x \in \{c, !c\}$, where $!c$ stand for not $c$, and $f(e, x)$ be a conversion function of an expression $e$ into the eliminated one. 

Here we write $\bar{G}$ for the eliminated grammar. Eliminating $c$-condition from $G$ is a conversion from $G$ into $\bar{G}$, and defined: for each production $A=e$ in $G$, two new productions $A_c$ and $A_{!c}$ are added into $\bar{G}$, as follows:

\[
A_c = f(e, c) ~~~~~ A_{!c} = f(e, !c)
\]

Now, we define the conversion function $f(e, x)$ recursively:

\begin{itemize}
\item $f(A,x)$ = $A_c$ if $x = c$, or $A_{!c}$ if $x = !c$
\item $f(e_1 e_2, x)$ = $f(e_1,x)  f(e_2,x)$
\item $f(e_1 / e_2,x)$ = $f(e_1,x) /  f(e_2,x)$
\item $f(\&e,x)$ = $\&f(e,x)$
\item $f(!e,x)$ = $! f(e,x)$
\item $f({\small \verb|<if|~ l ~ c \verb|>|},x)$ = $\epsilon$ if $x = c$, $!\epsilon$ if $x = !c$
\item $f({\small \verb|<if|~ l ~ !c \verb|>|},x)$ = $!\epsilon$ if $x = S_!c$, $\epsilon$ if $x = c$
\item $f({\small \verb|<on|~ c ~ e \verb|>|}, x)$ = $f(e, c)$
\item $f({\small \verb|<on|~ !c ~ e \verb|>|}, x)$ = $f(e, !c)$ 
\item $f(e, x)$ = $e$ if $e$ is none of the above
\end{itemize}

The number of productions in $\bar{G}$ is twice as many as that of $G$. That means that a grammar in which $n$ conditions are eliminated results in $O(2^n)$ productions in worst cases. 
In practice, we can make some unification for the same production such that $A_c = A_{!c}$, which may considerably reduce the number of productions in $\bar{G}$. Our empirical study (as described in Section \ref{sec:casestudies}) suggests that a single grammar involves not so many conditions that it would cause such an exponential increase of productions.

Note that the significant reason why we eliminate conditions is that condition operators may cause the serious invalidation of packrat parsing. In general, packrat parsing works on the assumption of stateless parsing\cite{ICFP02_PackratParsing}, while Nez adds new states such as ASTs, the symbol table, and parsing conditions. As we showed in Section \ref{sec:design}, the state of  ASTs does not influence any parsing behavior. The symbol table usually causes state changes with some character consumption, resulting in a fact that nonterminals rarely produce different results in the same position. As Grimm pointed out in \cite{PLDI06_Rats}, the {\em flow-forward} state change is not problematic in packrat parsing.  However, conditional parsing always causes state changes without any character consumption. This makes it easier to make a situation of different results of the same nonterminal in the same position. In fact, packrat parsing does not work without eliminating conditions from a Nez grammar. 

\section{Parser Runtime and Implementation} \label{sec:impl}

One of the advantages in open grammar is the freedom from parser implementation methods. In fact, the Nez parser tool, which we have developed with the Nez language, can generate three types of differently implemented parsers. First, as well as traditional parser generators, Nez produces parser source code written in the target language of the parser application. Second, Nez provides a grammar translator into other existing PEG-based grammars, (such as Rats$!$, PEGjs, or PEGTL). Third, Nez itself works as an efficient parser interpreter that loads a grammar at runtime. In this section, we briefly describe these three implementations. 

\subsection{Parser Runtime and Parser Generation} 

The parser runtime required in Nez is fundamentally lightweight and portable. Nez parsers, as well as PEG parsers, can be implemented with recursive descent parsing with backtracking. All productions are simply implemented with parse functions that compute on the input characters, and then nonterminal calls are computed by function calls. Backtracking is simply implemented over the call stacks. 

Notoriously, backtracking might cause an exponential time parsing in the worst cases, while packrat parsing\cite{ICFP02_PackratParsing} is well established to guarantee the linear time parsing with PEGs. The idea behind packrat parsing is a memoization whereby  all results of nonterminal calls are stored in the memoization table to avoid redundant nonterminal calls. Although the memoization table may require some complexity for its efficiency,  we use a simple and constant-memory memoization table, presented in \cite{PRO101}. 

In addition to the standard PEG parser runtime, Nez parsers require two additional state management to handle ASTs and symbols. Note that we assume that conditions are eliminated upfront, as described in Section \ref{sec:condition}. 

The AST construction runtime provides a parser with APIs that are based on AST operators in Nez. Importantly, Nez parsers are speculative paring in a way that some of the AST operations may be discarded when backtracking. To handle the consistency of ASTs, we support  transactional operations (such as {\sf commit} and {\sf abort}) and all operations are stored as logs to be committed. This transactional structure can be easily implemented with stack-based logging, and the AST construction is partially aborted at the backtracking time. Note that efficient packrat parsing with ASTs requires additional transactional management for ASTs. A detailed mechanism is reported in \cite{ASTMachine}.

The symbol table requires another state management. As with the operation logs in the AST constructor, we use a stack-based structure to control both the symbol scoping and backtracking consistency. The symbol table runtime provides a parser with APIs that enable adding symbols, eliminating symbols, and testing symbols to match the input.

Originally, we write the AST runtime and the symbol table runtime in Java. There is no use of functional data structure; instead, both operation logs and symbols are stored in a linked list, which is available in any programming language. Indeed, we have already ported Nez runtime into several languages, including C and JavaScript. The C version of the AST runtime is at most 500 lines in code and the symbol table runtime is at most 200 lines in code, suggesting its high portability. 

\subsection{Grammar Translation}

A Nez grammar is specified with a declarative form of the AST operators and symbol operators, which are performed as a kind of action in the parsing context. These actions are limited in number and can be statically translated into code fragments written in any programming language. This means that a Nez grammar is convertible into PEGs 
with embedded semantic action code.  

Grammar translation is another approach to the Nez parser implementation. The advantage is that we enjoy optimized parsers that existing PEG tools generate.  On the other hand, impedance mismatch occurs between two grammars. In particular, many existing PEG-based parser generators have their own supports for AST representations. For example, PEGjs\cite{PEGjs} produces a JSON object as a semantic value of all nonterminal calls. In those cases, the translation of AST operators is unnecessary in practice. The symbol operators are always translatable if the symbol runtime that run on a host language is readily available.  

Figure \ref{fig:action} shows an example of converting symbol operators into  semantic actions \verb| { ... } | and semantic predicates \verb| &{ ... } |. The rollback of the symbol table at the backtracking time is automated before attempting alternatives. As shown, a Nez grammar is recursively convertible into PEGs with action code using the Nez runtime.

\begin{figure}[tb]
\begin{small}
\begin{tabular}{ll}
$e~/~e'$ & $\verb|{symbolTable.startBlock()}|$ \\
 & $ e ~ \verb|{ symbolTable.commit() } | $  \\
& $ / ~ \verb|{symbolTable.abort()} | ~  e'$  \\

$\verb|<symbol|~A\verb|>|$ &
$\verb|{ symbolTable.add(A, capture(|A\verb|)) }|$ \\

$\verb|<block |~e\verb|>|$ &
$\verb|{ symbolTable.startBlock() }|$ \\
 & $ e $ \\
 & $\verb|{ symbolTable.abortBlock() }|$ \\

$\verb|<local |~A~e\verb|>|$ &
$\verb|{ symbolTable.startBlock().mask(A) }|$ \\
 & $ e $ \\
 & $\verb|{ symbolTable.abortBlock() }|$ \\

$\verb|<exists|~A\verb|>|$ &
$\verb|&{ symbolTable.count(A) > 0 }|$ \\

$\verb|<exists|~A~x\verb|>|$ &
$\verb|&{ symbolTable.top(A) == x }|$ \\

$\verb|<match|~A\verb|>|$ &
$\verb|&{ match(symbolTable.top(A)) }|$ \\

$\verb|<is|~A\verb|>|$ &
$\verb|&{ symbolTable.top(A) == capture(A) }|$ \\

$\verb|<isa |~l\verb|>|$ &
$\verb|&{ symbolTable.contains(A, capture(A)) }|$ \\

\end{tabular}
\end{small}

\caption{Implementing Nez symbol operators with action code}
\label{fig:action}
\end{figure}

Currently, the Nez tool provides the grammar translation into PEGjs, PEGTL, and LPeg, although the translation of the AST construction might be restricted due to the impedance mismatch. Note that translating a Nez grammar into a LALR or LL grammar is an interesting challenge, though beyond the scope of this paper.

\subsection{Virtual Parsing Machine}

The simplicity of PEGs makes it easier to achieve dynamic parsing in a way that a parser interpreter loads a grammar  at runtime to parse the input. Since dynamic parsing requires no source compilation process, dynamic parsing makes it easier to import the Nez parser functionality as a parser library into parser applications. We consider that dynamic parsing is better suitable for many use cases of open grammars. Accordingly, the standard Nez parser is based on dynamic parsing and then implemented on top of a virtual parsing machine, an efficient implementation of dynamic parsing. 

A Nez parsing machine is a stack-based virtual machine that runs with a set of bytecode instructions, specialized for PEG, AST, symbol, and memoization operators. Table \ref{table:insts} is a summary of the bytecode instructions. A parsing expression $e$ in Nez is converted to bytecode by using a compile function $\tau(e, L)$, where $L$ is a label representing the next code point. Figure \ref{fig:compile} shows an excerpt of the definition of $\tau(e, L)$. The byte compilation is simply inductive, while Nez supports additional super instructions to generate optimized bytecode.  

Currently, the Nez parser tool includes a bytecode compiler and a parsing machine, both written in Java. In addition, the parsing machine is ported into C and JavaScript. Especially, the C version of the Nez parsing machine is highly optimized with indirect thread code and SEE4 instructions. The performance evaluation is studied in Section \ref{sec:perf}.

\begin{table}[tb]
\begin{tabular}{lp{6cm}}
PEG &
\textsf{
\noindent
nop fail alt succ jump call ret pos back skip byte any 
} \\
AST &
\textsf{
\noindent
tpush tpop tleftfold tnew tcapture ttag treplace 
tstart tcommit tabort 
} \\
Symbol &
\textsf{
\noindent
sopen sclose smask symbol exists isdef match is isa 
} \\
Memo &
\textsf{
\noindent
lookup memo memofail tlookup tmemo
} \\

\end{tabular}

\caption{Instructions of the Nez parsing machine}
\label{table:insts}

\end{table}

\begin{figure}[tb]

\begin{small}
\begin{center}

\begin{tabular}{lrrl}
$\tau(A = e,\: L)$      &  \verb|      = | & $A$ & $\tau(e,\: L')$  \\
&  & $L'$ & $\kw{ret}$  \\
&  &  &   \\
$\tau(\epsilon,\: L)$      &  \verb|      = | & & $\kw{nop}$  \\
$\tau(a,\: L)$      &  \verb|      = | & & $\kw{byte} ~a$  \\
$\tau(A,\: L)$      &  \verb|      = | & & $\kw{call}~ A$  \\

$\tau(e_1 ~ e_2,\: L)$ &  \verb|      = | & & $\tau(e_1,\: L')$  \\
 &  & $L'$ & $\tau(e_2,\: L)$  \\

$\tau(e_1  /  e_2,\: L)$ &  \verb|      = | & & $\kw{alt} ~L'$  \\
& & & $\tau(e_1,\: L)$  \\
&  & & $\kw{succ}$  \\
&  & $L'$ & $\tau(e_2,\: L)$  \\

$\tau(\&e,\: L)$ &  \verb|      = | & & $\kw{pos}$  \\
& & & $\tau(e,\: L')$  \\
&  & $L'$ & $\kw{back}$  \\

$\tau(!e,\: L)$ &  \verb|      = | & & $\kw{alt} ~L$  \\
& & & $\tau(e,\: L')$  \\
&  & $L'$  & $\kw{succ}$  \\
&  &  & $\kw{fail}$  \\ \hline

$\tau(\{e\},\: L)$ &  \verb|      = | & & $\kw{tnew}$  \\
& & & $\tau(e,\: L')$  \\
&  & $L'$  & $\kw{tcapture}$  \\

$\tau(\$(e),\: L)$ &  \verb|      = | & & $\kw{tpush}$  \\
& & & $\tau(e,\: L')$  \\
&  & $L'$  & $\kw{tlink}$  \\
&  &   & $\kw{tpop}$  \\

$\tau(\#t,\: L)$ &  \verb|      = | & & $\kw{ttag}~\#t$  \\ \hline

$\tau(\langle block~e\rangle,\: L)$ &  \verb|      = | & & $\kw{sopen}$  \\
& & & $\tau(e,\: L')$  \\
&  & $L'$  & $\kw{sclose}$  \\

$\tau(\langle local~A~e\rangle,\: L)$ &  \verb|      = | & & $\kw{sopen}$  \\
& & & $\kw{mask}~A$  \\
& & & $\tau(e,\: L')$  \\
&  & $L'$  & $\kw{sclose}$  \\

$\tau(\langle symbol~A\rangle,\: L)$ &  \verb|      = | & & $\kw{pos}$  \\
& & & $\tau(A,\: L')$  \\
&  & $L'$  & $\kw{symbol}$  \\

$\tau(\langle exists~A\rangle,\: L)$ &  \verb|      = | & & $\kw{exists}~A$  \\
$\tau(\langle match~A\rangle,\: L)$ &  \verb|      = | & & $\kw{match}~A$  \\

$\tau(\langle is~A\rangle,\: L)$ &  \verb|      = | & & $\kw{pos}$  \\
& & & $\tau(A,\: L')$  \\
&  & $L'$  & $\kw{is}~A$  \\
\end{tabular}

\end{center}

\end{small}

\caption{Definition of a compile function of $e$ with core instructions}
\label{fig:compile}

\end{figure}

\section{Case Studies and Experiences} \label{sec:casestudies}

We have developed many Nez grammars, ranging from programming languages to data formats. All developed grammars are available online at \url{http://github.com/nez-peg/nez-grammar}. 
Table \ref{table:grammars} shows a summary of the major developed grammars.The column labeled "\#G" indicates the number of {\em defined} productions in a grammar, while the column labeled "\#P" indicates the number of {\em parser} productions through eliminating the conditional operators. 
The same numbers of productions in \#G and \#P means no use of conditional parsing. 
The column labeled "Symbols and Conditions" indicates nonterminals and conditions that are used in the grammar.

\begin{table}[tb]
\begin{tabular}{lrrl}
Language & \#G  & \#P & {\tt Symbols} and {\it Conditions} \\ \hline
C & 102 & 102 &   {\tt TypeName} \\
C\#5.0 & 897 & 1325 &  {\it AwaitKeyword}, {\it Global} \\
CoffeeScript & 121 & 121 &  {\tt Indent} \\
Java8 & 185 & 185 & \\
JavaScript & 153 & 166 & {\it InOperator} \\
Konoha & 163  & 190 & {\it Offside} \\
Lua5 & 102 & 105 & {\it MultiLineBracketString} \\
Python3 & 139 & 155 & {\tt Indent}, {\it Offside} \\
Ruby & 420 & 580 & {\small {\tt Delimiter}, {\it Primary}, {\it DoExpr}} \\
\hline

\end{tabular}
\caption{Grammar summary: the {\tt typewriter} font stands for a symbol name and the {\it italic} font stands for a condition name}
\label{table:grammars}
\end{table}

Here are some short comments on each of the interesting developed grammars. 

\begin{itemize}

\item C -- Based on two PEG grammars written in Mouse. The semantic code embedded to handle the {\tt typedef}-defined name is converted into the {\tt TypeDef} symbol. The nested typedef statements are not implemented.

\item C\# -- Developed from scratch, referencing C\#5.0 Language Specification (written in a natural language). The condition {\it AwaitKeyword} is used to express contextual keyword {\tt await}. 

\item Java8 -- Ported from Java8 grammar written in ANTLR4\footnote{https://github.com/antlr/grammars-v4/blob/master/java8/Java8.g4}. Java can be specified without any Nez extensions. 

\item JavaScript -- Based on JavaScript grammar for PEG.js\footnote{https://github.com/pegjs/pegjs/blob/master/examples/javascript.pegjs}. We simplify the grammar specification using the parsing condition {\it InOperator} that distinguishes expression rules from the {\tt for/in} context. 

\item Konoha -- Developed from scratch. Konoha is a statically typed scripting language, which we have designed in \cite{Konoha}. The most syntactic constructs come from Java, but we use the parsing condition {\tt SemicolonInsertion} for expressing the conditional end of a statement. More importantly, Konoha is a language implementation that uses  ASTs constructed by the Nez parser.

\item Python3 -- Ported from Python 3.0 abstract grammar\footnote{https://docs.python.org/3.0/library/ast.html}. The {\tt Indent} is used to express the indentation-based code block by capturing white spaces.  

\item Ruby -- Developed from scratch. The {\tt Delimiter} is used to express the context-sensitive delimiting identifier in the Here document.

\item Haskell -- Postponed. We didn't express the indent symbol for Haskell's code layout since the indentation starts in the middle of expressions. This is mainly because of a limitation of a symbol operator in Nez; if Nez provided Haskell-specialized symbol operator such as \verb|<haskell-indent>|, we could handle the indentation-based code layout through the symbol tables. 

\end{itemize}


Two findings are confirmed throughout our case studies.  First, many of the programming languages, as listed in Table \ref{table:grammars}, can not be well expressed with pure PEG. The introduction of symbol tables significantly improve the expressiveness, although some corner cases might remain. Even in corner cases, we consider that the symbol table approach is still applicable.  
Second, the parsing condition, although it does not directly improve the expressiveness of PEGs, simplifies the specification task. The growth of \#D/\#G indicates that the grammar developer would specify as many productions  as indicated at \#G if the conditional parsing were not supported in Nez. Furthermore detailed discussions and earlier reports on grammar developments with examples are available in \cite{JIPSE2015}. 

All grammars we have developed include the AST construction. 
At the moment, the only working example of a language implementation with Nez is with Konoha, a reimplemented version of \cite{Konoha} as a JVM-based language implementation, including type checking and code generation on top of the ASTs that are constructed by a Nez parser. 
Since each AST node records the source location when capturing a string, 
Konoha can carry out error reporting, although it is very primitive. 


\begin{figure*}[t]
\begin{center}
\includegraphics[width=16cm]{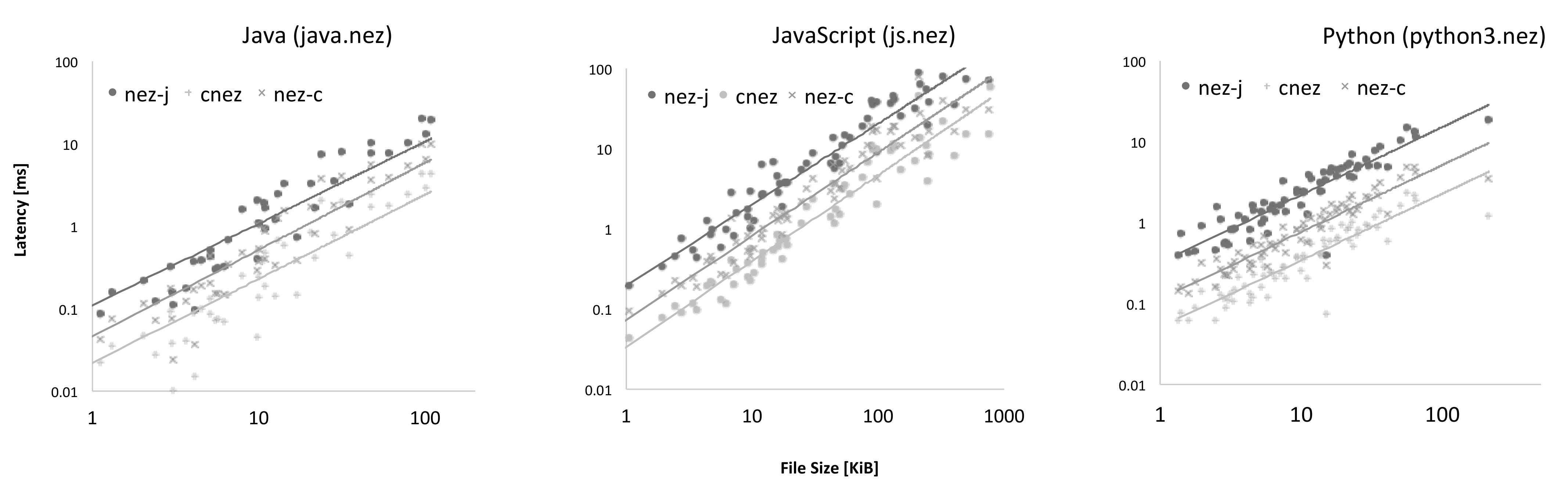}
\caption{Parsing time of the open source files for Java, JavaScript, and Python against the input size plotted as log-log base 10. }
\label{fig:linear}
\end{center}
\end{figure*}

\begin{figure}[tb]
\begin{center}
\includegraphics[width=8cm]{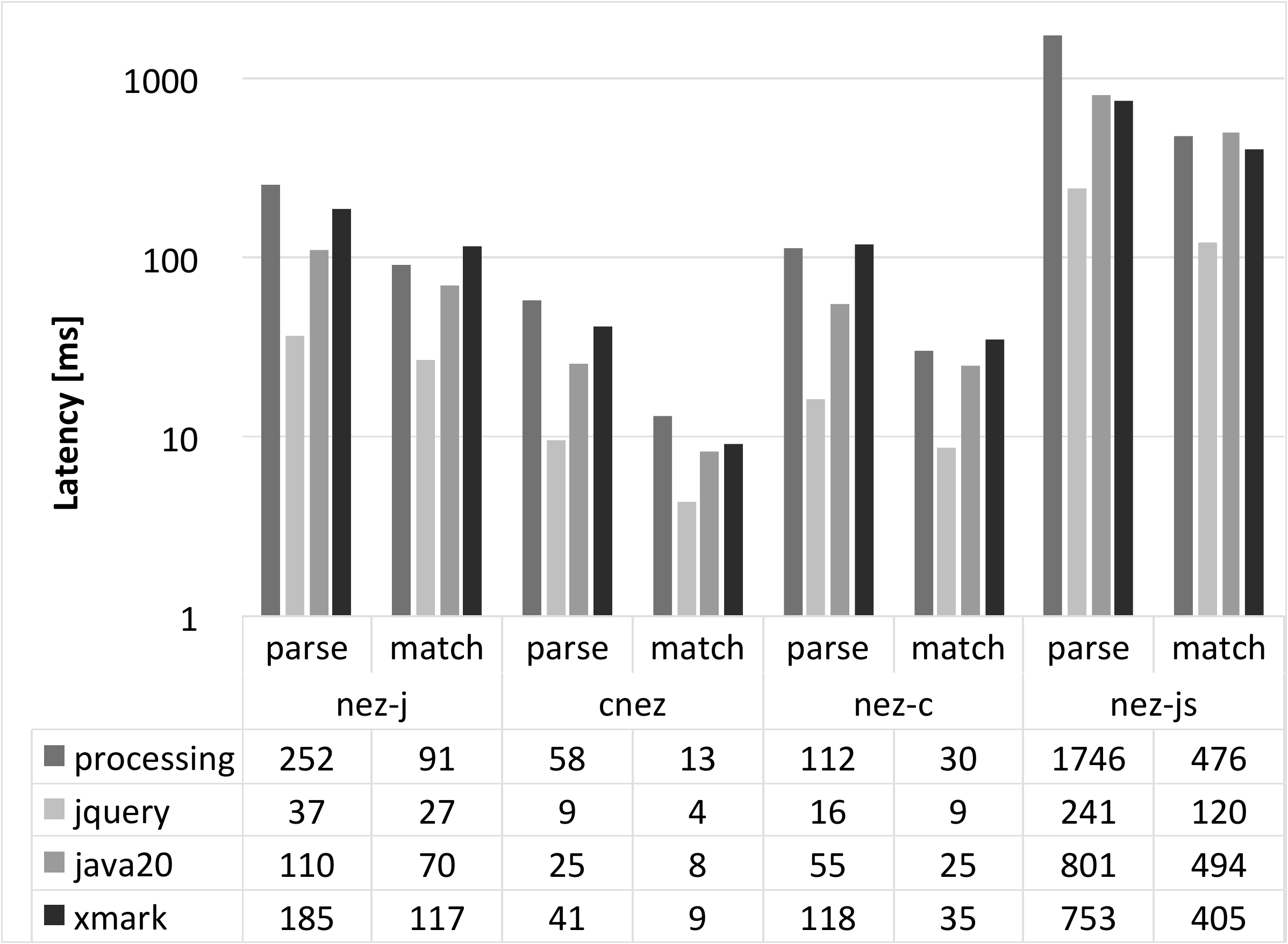}
\caption{Performance comparison of differently implemented Nez parsers: {\sf nez-j}, {\sf nez-c}, {\sf cnez}, and {\sf nez-js}}
\label{fig:nezperf}
\end{center}
\end{figure}

\begin{figure*}[tb]
\begin{center}
\includegraphics[width=14cm]{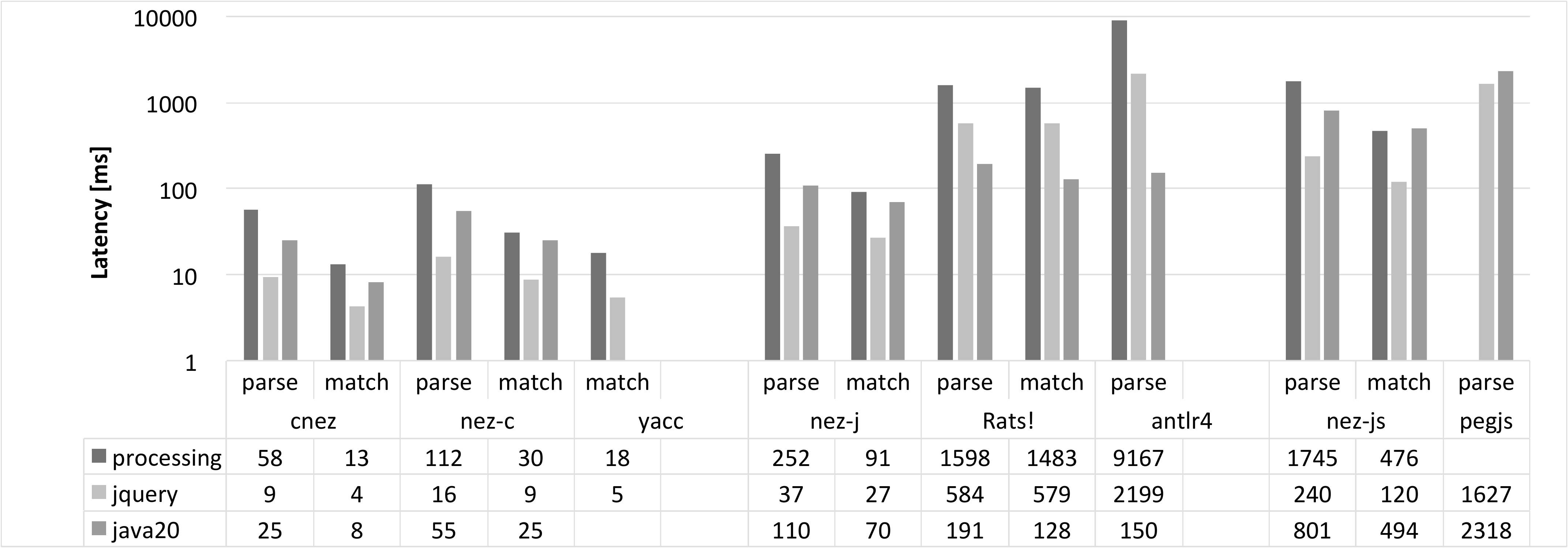}
\caption{Performance comparison of {\sf processing}, {\sf jquery}, and {\sf java20} in various parsers. The {\sf pegjs} parser cannot end parsing {\sf process} due to its out of memory errors. }
\label{fig:mainperf}
\end{center}
\end{figure*}

\section{Performance Study} \label{sec:perf}

This section is an experimental report on our performance and comparative study. We start by describing common setups. 
All tests in this paper are measured on DELL XPS-8700, with 3.4GHz Intel Core i7-4770, 8GB of DDR3 RAM, and running on Linux Ubuntu 14.04.1 LTS. 
All C programs are compiled with clang-3.6, all Java parsers run on Oracle JDK 1.8, and all JavaScript parsers run on node.js version 5.0. 
Tests are run several times, and we record the average time for each iteration. 
The execution time is measured in millisecond by {\small {\tt System.nanoTime()}} in Java APIs and {\small {\tt gettimeofday()}} in Linux. Tested source files are randomly collected from major open source repositories. To improve the time accuracy, we eliminate small files from the data set. 

All tested Nez grammars are available online as described in Section \ref{sec:casestudies}. For convenience, we label tested grammars: {\sf java.nez}, {\sf js.nez}, {\sf python3.nez}, {\sf xml.nez}, and {\sf xmlsym.nez}. Note that {\sf xmlsym.nez} is a symbol version of {\sf xml.nez}, which checks whether the closed tag is equal to the open tag, as described in Section 2. The two tested grammars {\sf python3.nez} and {\sf xmlsym.nez} contain symbol operators in a grammar. 

Using Nez grammars, we generate Nez parsers for Java, C, and JavaScript. The generated parsers are grouped by the base implementation with the following labels:

\begin{itemize}
\item {\sf nez-j} - an original Nez parsing machine on JVM, incorporated as a standard Nez parser,
\item {\sf nez-c} - a C-ported Nez parsing machine, 
\item {\sf nez-js} - a JavaScript-ported Nez parsing machine, and
\item {\sf cnez} - generated C parsers from the Nez tool
\end{itemize}

Linear time paring is a primary concern of parser implementations. 
Figure \ref{fig:linear} shows the parsing time plotted against file sizes in, respectively, Java, JavaScript, and Python.
The lines are the linear regression line fit of {\sf nez-j}, {\sf cnez}, and {\sf nez-c} parsers, suggesting the paring time is linear. Since Nez parsers are integrated with packrat parsing,  we have not observed any super-linear parser behavior throughout this experiment.

\begin{figure}[tb]
\begin{center}
\includegraphics[width=6.5cm]{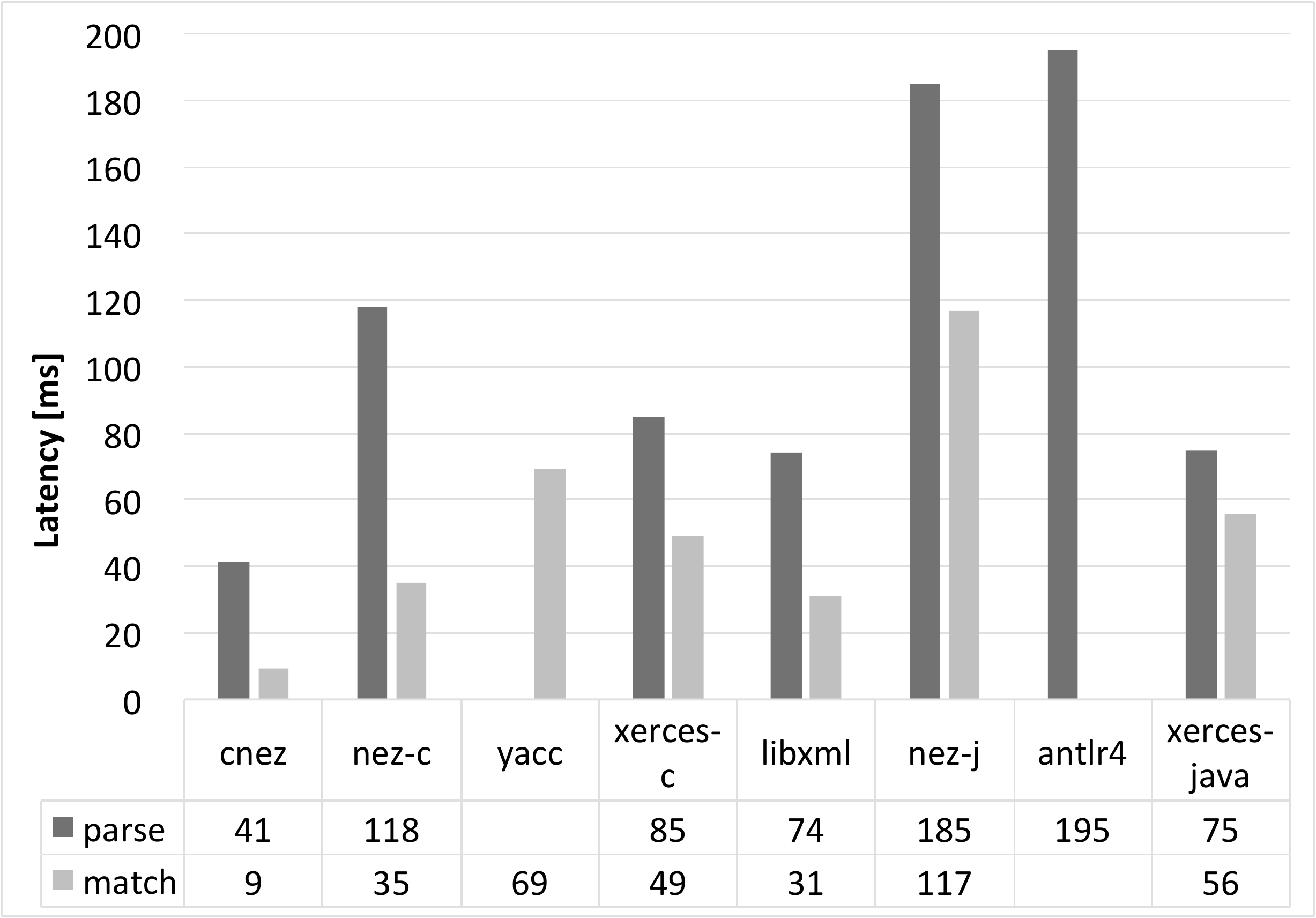}
\caption{Parsing time in {\sf xmark}}
\label{fig:xmlperf}
\end{center}
\end{figure}

\begin{figure}[tb]
\begin{center}
\includegraphics[width=6.5cm]{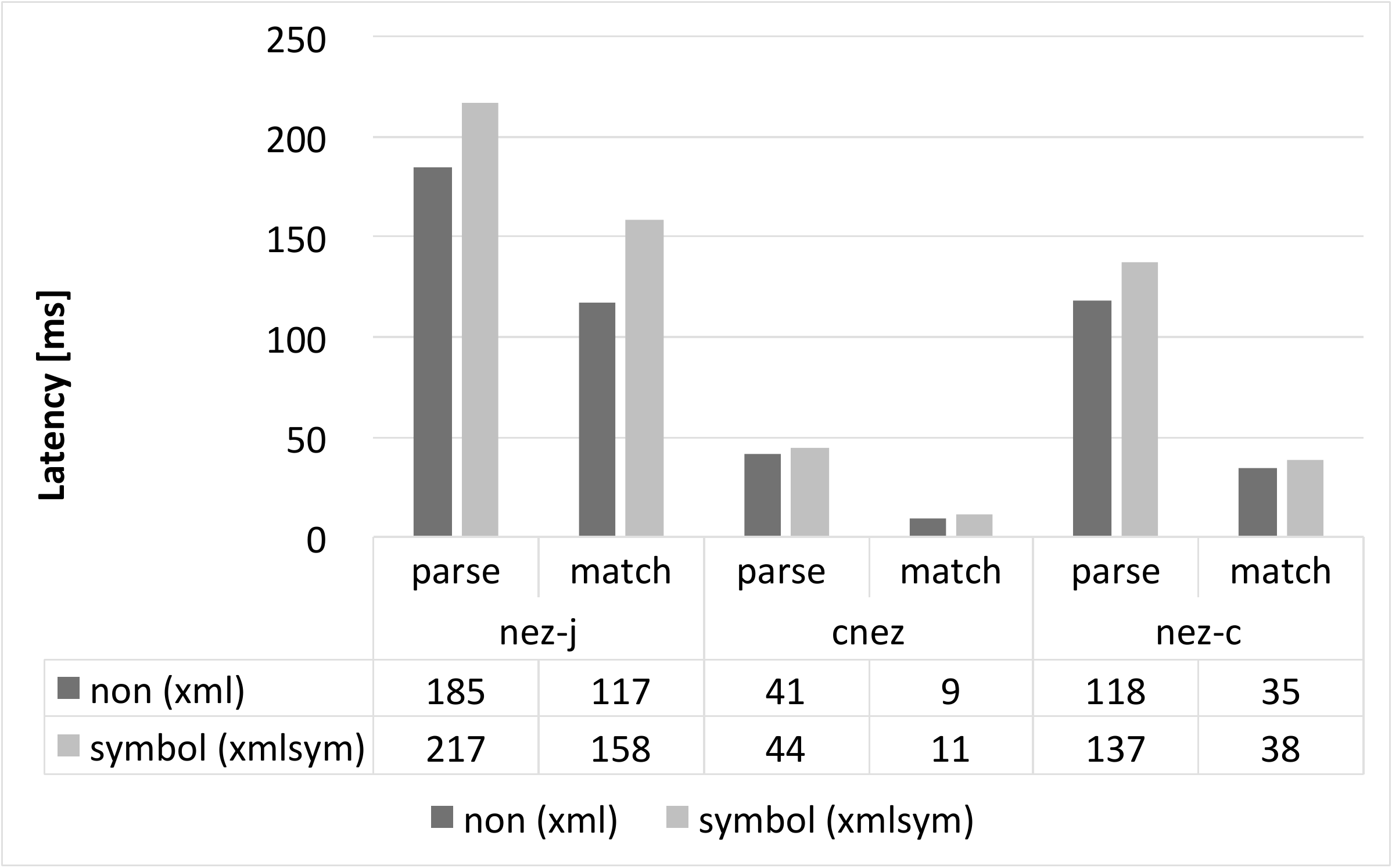}
\caption{Effects of symbol operators }
\label{fig:symperf}
\end{center}
\end{figure}

To make detailed analysis, we have chosen some of tested source files from the collection. The chosen data are labeled as:

\begin{itemize}
\item {\sf processing} -- a JavaScript version of the Processing runtime
\item {\sf jquery} -- a very popular JavaScript library. We used a compressed source.
\item {\sf java20} -- 20 Java files collection, consisting of relatively large files (more than 10KB size). The parsing time is recorded as the summation of all 20 parsing times. 
\item {\sf xmark} --  a synthetic and scalable XML files that are provided by XMark benchmark program \cite{VLDB02_XMark}. We used a newly generated file in 10MB. 
\end{itemize}

Figure \ref{fig:nezperf} shows the parse time of Nez parsers. The data point {\tt parse} indicates the parsing time including the AST construction, while {\tt match} indicates the parsing time without the AST construction. The time scale is plotted as log base 10 due to the variety of results. 
The {\sf cnez} parsers are fastest as easily predicted. In comparisons among the parsing machines, the {\sf nez-c} parser is almost 2.5 times faster than {\sf nez-java}, and almost 10 times faster than the {\sf nez-js}.  

Next, we will turn to the performance comparison with existing standard parsers. Here, we have chosen Yacc\cite{Yacc}, ANTLR4\cite{OOPSLA14_Antlr}, Rats$!$\cite{PLDI06_Rats}, and PEGjs\cite{PEGjs}, since they notably produce efficient parsers and are accepted in many serious projects. We set up these parser generators as follows.

\begin{itemize}

\item {\sf yacc} -- a standard LALR parser generator for C. In this experiment, we suffered from the limited availability of grammars, and could only run a JavaScript parser without AST construction. 

\item {\sf antlr4} -- a standard parser generator based on the extended LL($k$) with some packrat parsing feature. All tested grammars are derived from the official ANTLR4 grammar repository.

\item {\sf rats$!$} -- a PEG-based parser generator\cite{PLDI06_Rats} for Java. Java grammar is derived from a part of the xtc implementation, while JavaScript parser is generated from a Nez grammar {\sf js.nez} by the grammar translation.  

\item {\sf pegjs} -- a PEG-based parser generator\cite{PEGjs} for JavaScript. JavaScript grammar is derived from its official site, and Java grammar is translated from {\sf java.nez} by Nez tools.

\end{itemize}

Figure \ref{fig:mainperf} shows the parsing time of {\sf processing}, {\sf jquery}, and {\sf java20} in each of the examined parsers. All Nez parsers show better performance with competitive parsers in each of the host environments. In ANTLR4 and Rats$!$, some of results are surprisingly bad. We run fairly in terms of the time measurement and the JIT condition, but we are still not sure that Rats$!$ and ANTLR4 parsers are best optimized in the experiment. As a result, we conclude that Nez parsers achieve {\it competitive} performance compared to these practical parser generators. Likewise, we compare Nez parsers for {\sf xmark} with standard XML parsers such as libxml and xerces, which are highly optimized by hand. Figure \ref{fig:xmlperf} shows the comparison of parsing time in {\sf xmark}. The data points {\sf match} and {\sf parse} in XML parsers correspond to SAX and DOM parsers, respectively. Although we have observed the overhead of AST construction in Nez parsers, they are also competitive. 

Finally, we focus on the performance effect of symbol operators. Figure \ref{fig:symperf} shows the performance comparison of {\sf xml.nez} and {\sf xmlsym.nez}. We confirm that the costs of symbol operators, requiring \verb|<block>|, \verb|<symbol>|, and \verb|<is>| at each closed tag, are small at the acceptable level. 

\section{Related Work} \label{sec:relatedwork}

Developing parsers is ubiquitous in many applications, and hand-written parsers are obviously prone to errors. 
Parser generation from a formal grammar specification has a long history in programming language research. A common key challenge is balancing the expressiveness of formal grammars and how to implement them in practical manners. In the early 1970s, Stephen C. Johnson developed Yacc\cite{Yacc}, based on a LALR(1) grammar with embedded C code to perfom user-defined actions. Since Yacc successfully hits a sweet spot between the expressiveness and practical parser generation, the use of action code has been broadly adopted in most of the major parser generators \cite{PLDI11_Antlr,OOPSLA10_ParadiseLost,OOPSLA14_ParserCombinator} across various grammar foundations, including LL($k$), GLR, and PEG.

In parallel, the problems of action code have been pointed out, for example, the lack of grammar reuse\cite{ICPC08_SemanticActions}, decreased maintainability \cite{LDTA10_DSL}, and ad hoc behavior\cite{POPL13_Indentation}. There have been several attempts to decrease the disadvantage of action code. 
Terence Parr, the author of ANTLR, has proposed a prototype grammar, inspired by the revision control system\cite{ICPC08_SemanticActions}. Elkhound\cite{CC04_Elkhound}, for instance, allows the users to write action code written in multiple programming languages. 
Despite these attempts, grammar reuse is still limited and arbitrary action inevitably requires porting to other languages. 

The idea of open grammars has been strongly inspired by the article: {\em Pure and declarative syntax definition: paradise lost and regained}\cite{OOPSLA10_ParadiseLost}. SDF+ADF and Stratego\cite{MetaBorg,SCP08_Stratego} intend their users to write a syntactic analysis using an algebraic transformation. The actions are not arbitrary, but  they are expressive enough. However, the actions differ from those required in Nez, since SDF is based on Generalized LR\cite{PHD97_SGLR}, where actions are mainly used to resolve grammar ambiguity. Nez, on the other hand, is unambiguous as a PEG, and symbol operators are used for handling state changes. 

Data-dependent grammars \cite{POPL10_Yakker,ONWARD15_Iguana} share similar ideas in terms of recognizing context-sensitive syntax. Briefly, these grammars uses a variable to be bound to a parsed result, which seems to be the addition of symbols in Nez. Data-dependent grammars are more expressive in a way that they express more complex condition such as \verb|([n>0] OCTET{n:=n−1})|, while Nez is more declarative and leads to better readability. In addition, the better expressiveness of data-dependent grammars suggests a future promising extension of Nez, including numerical values in the symbol table.

Among many formal grammars such as LALR($k$) and LL($k$), PEGs are simple and seemingly a suitable foundation for implementing portable parser runtime. In the reminder of this section, we would like to focus on PEG-specific related work.

{\bf Extensions.} 
With Rats$!$, Robert Grimm has shown an elegant integration of PEGs with action code despite its speculative parsing\cite{PLDI06_Rats}. Many other PEG-based parser generators\cite{PEGjs,FI07_Mouse,PEGTL} have adopted the semantic action approach for expressing syntax constructs that PEGs can not recognize. Yet, there have been several declarative attempts for PEG extensions.
Adams has recently formalized  Indent-Sensitive CFGs \cite{POPL13_Indentation} for the indentation-based code layout and has extended the PEG-version of IS-CFGs \cite{Haskell14_Indentation}.
Iwama et al have extended PEGs with the $e~U~e$ operator to combine a black-boxed parser for a natural language\cite{ICSE12_PEGs}. Compared to these existing extensions, Nez more broadly covers various parsing aspects.

{\bf AST Construction}. 
Traditionally, a grammar developer usually writes action code to construct some forms of ASTs as a semantic value\cite{LDTA10_ASTConstruction}. However, writing the AST construction is tedious, and many parser generators have some annotation-based supports to automate the AST construction. The underlying idea is to filter trees that are derived from structural parse results of nonterminal calls. However, the filtering approach is problematic in PEGs since PEGs disallows the left-recursive structure of derived trees. Recently, handling the left recursion in PEGs has been established in \cite{SCP14_Left}, although another annotation is needed for preserving operator precedence. 
Nez allows users to specify an explicit structure of ASTs (including tags and labels), resulting in a more general string-to-tree transducer. Notably, a similar capturing idea appears in LPeg\cite{LPeg}, a PEG-based pattern matching, but Nez enables more structural complexity for capturing structured data. 

{\bf Parser Runtime}.
Nez parser runtime is based on many of previous works reported in the literature\cite{PLDI06_Rats,DLS08_LPeg,FI07_Mouse,PRO101}. In particular, the state management in speculative parsing is built on the partial transaction management of Rats$!$.  The Nez virtual machine is said to be an extended version of the LPeg machine. Notably, the Nez machine has instruction supports for packrat parsing \cite{ICFP02_PackratParsing}, which is a major lack in the LPeg machine. The originality of Nez parsers is that we combine these established techniques in a way that the parser runtime is still simple and portable.  


\section{Conclusion} \label{sec:conclusion}

Nez is a simple, portable, and declarative grammar specification language. As an alternative to action code, Nez provides a small set of extended PEG operators that allow the grammar developer to make AST constructions, as well as context-sensitive and conditional parsing. Due to these extensions, Nez can recognize a language that cannot be well expressed by pure PEGs. In addition, the Nez parser runtime is also simple and portable. Based on the Java-based implementation, we have developed C and JavaScript ports, suggesting that the Nez parser runtime is implementable in any modern programming languages. More importantly, Nez parsers achieve practical and competitive performance in each of the ported environments.

This paper is an initial report on Nez and the idea of open grammars. Further evaluations are obviously necessary on the expressiveness of Nez with extensive case studies. To achieve the aim of open grammars, a lot of interesting challenges remain. Future work that we will investigate includes grammar modularity and inference, a tree checker for better connectability of parser applications, and a DFA-based  machine for more efficient parsing. In the end, we hope that Nez parsers will be available in many programming languages. Our developed tools and grammars are available online at \url{http://nez-peg.github.io/}. 
   
\acks
The C version of Nez parsers (cnez and virtual parsing machines) have been developed by Masahiro Ide and Shun Honda.The JavaScript version has been developed by Takeru Sudo. Finally, the anonymous reviewers are acknowledged for their helpful suggestions for improving an earlier draft of this paper.


\bibliographystyle{abbrvnat}
\bibliography{../bib/parser,../bib/mypaper,../bib/url,../bib/data}  

\end{document}